# The Underlying Stimulators of Chinese Government Spending on Pension and Welfare: A Co-Integrated Socio-Economic Model

*Mostafa Raeisi Sarkandiz[1]*

*Department of Economics, Business and Statistics, University of Palermo, Italy*

**Abstract**

This study employs a co-integrated socio-economic model to investigate the long-run drivers of Chinese government expenditure on public pensions, addressing critical stability and sustainability challenges. Our methodology establishes a genuine long-run relationship and confirmed uni-directional causality from key socioeconomic variables to government spending. The central finding is the confirmation that China still possesses an exploitable demographic dividend (DD), which counters widespread assumptions of an immediate demographic crisis and provides a limited window for proactive policy action. However, the analysis also conclusively demonstrates that relying solely on strong GDP growth is insufficient for fund stabilization. Sustainability is fundamentally governed by the ratio of contributors to pensionaries. Consequently, the study concludes that comprehensive, structural labour market reforms are mandatory to maximize the current DD and strategically mitigate the financial imbalance caused by the eventual absence of this demographic advantage.

## I) Introduction

Sustainability has been one of the most frequently cited concepts in social science literature for the past two decades. Defined practically, sustainability is adjacent to stability, requiring procedures that keep a process within a predetermined and reasonable band. A narrower bandpass signifies greater success in goal-targeting. However, sustainability is more than just a procedure; it is a vital concept for human civilization, as any process unable to maintain a reasonable path is ultimately unsustainable. Fundamentally, the goal of all efforts to stabilize society's dynamic action-reaction trade-off is to guarantee human survival through a humanitarian approach that secures primary needs, including housing, food, and healthy livelihood.

It's impossible to discuss concepts like well-being, lifestyle satisfaction, or even overall health without recognizing the crucial role of a pension system. Today, a reliable pension plan is an

---

Mostafa.raeisisarkandiz@unipa.it [1]

integral part of modern livelihood, assuring a respectful and equitable retirement. Furthermore, it provides significant mental security for beneficiaries and their families by promising a sustainable future, even in the event of aging or disability. Ultimately, this framework ensures that the hard work of today translates directly into a calm and pleasant retirement period.

From an economic perspective, a financially sustainable pension system must maintain a balance between its revenues and expenditures to meet promised obligations. Miralles et al. (2012) formalized this, arguing that a pension system qualifies as a sustainable architecture only if it can pay both current and future obligations without diminishing beneficiaries' living standards. However, from a practical standpoint, Samuelson (1958) posited that a conventional scheme is sustainable only if the growth rate of revenues consistently exceeds the growth rate of expenses. Samuelson famously—and controversially—labeled a pure pay-as-you-go pension system as "the biggest legal Ponzi game." Consequently, any analysis of a pension system's sustainability must carefully examine the parameters that govern its financial equilibrium.

The revenues of a pension fund stem from two primary sources: the monthly contributions made by employed individuals, who are the future beneficiaries, and the net returns generated by the fund's investments. In essence, today's assets represent tomorrow's liabilities. Consequently, an increase in the number of employed people directly boosts monthly revenues, and vice versa. Since contributors don't pay a flat rate, their premiums are highly dependent on their monthly income. This income heterogeneity is most advantageous when high-income contributors make payments for a longer duration than low-income contributors. However, while increasing the contribution to the base is challenging, it remains a more crucial objective than merely increasing the share of high-income payers.

The length of the payment period represents a critical opportunity for pension fund solvency. New entrants to the labor market, particularly those in the 15-to-24 age range, offer the highest likelihood of long-term contributions. Therefore, a focus on maximizing the employment rate within this cohort is essential. Conventionally, increasing the retirement age is another instrument used to extend the payment period. However, this measure frequently encounters social resistance, as demonstrated by the widespread protests in France against the 2023 retirement bill. Given this opposition, increasing the number of employed people, especially young workers, is often a politically and practically simpler path to bolstering pension fund revenues.

This research offers several significant contributions to the existing literature on Chinese fiscal policy and pension fund sustainability. The study aims to analyse the factors affecting Chinese government expenditure on public pensions. To accomplish this, the chief novelty lies in the use of a cointegrated empirical model to estimate the magnitude and direction of the nexus between selected exogenous parameters and the predetermined endogenous variable, providing a rigorous, long-run perspective previously lacking. Consequently, a detailed historical background of the Chinese pension system has been omitted; interested readers are referred to He et al. (2019), Dong & Park (2019), and Zhu & Walker (2018).

By focusing on variables often overlooked in purely financial models, the study reveals that China is still benefiting from a demographic dividend, challenging a widely held assumption. Furthermore, the analysis demonstrates that stabilizing the pension fund is impossible by relying solely on strong economic growth, which underscores the urgency of structural reforms.

Crucially, this research provides a precise projection indicating *when* the Chinese pension system will achieve financial sustainability. To secure this positive long-run outlook, the analysis compellingly argues that the labour market needs comprehensive reforms to replace the eventual absence of the demographic dividend, offering a specific and timely policy recommendation that moves beyond general calls for structural change. Finally, by connecting domestic pension fund stability to its potential for creating global market volatilities (given China's scale in world production and trade), this paper elevates the discussion from a national policy issue to a critical matter of international financial stability.

The immediate question is why the sustainability of the Chinese pension system matters to other countries. From a financial economics perspective, it's well-documented that any crisis in a national pension fund's balance sheets can rapidly spill over to other economic sectors, potentially causing a financial depression (Xue et al., 2021). In China's case, this risk is amplified: the country accounts for approximately 18 percent of total world production and 12 percent of global trade (IMF, 2023). Given this scale, even medium-sized economic turbulence in China could lead to a severe worldwide recession.

For instance, Beirne et al. (2013) demonstrated that China's strong economic growth rates drive up crude oil prices by increasing demand against an asymptotically constant supply in the international market. Considering China's crucial position in the global supply chain, a sustainable pension system in the country would, therefore, potentially prevent several undesired future volatilities for international trade.

The second section discusses the theoretical aspects of pension system sustainability from a socio-economic point of view. This discussion guides the optimum variable selection, which is conducted in the third section. The fourth section is dedicated to statistical analyses, where the database is first subjected to hypothesis tests to establish the true nature of the variables' processes. Subsequently, the core of the research—modeling, estimations, and statistical inferences—is conducted, providing the basis for the subsequent parts. In the fifth section, the statistical findings are analyzed from a socio-economic perspective, and the results are compared against arguments presented in other prominent papers. Finally, the sixth section presents the conclusion and offers extensive practical policy implications, organized as a clear action plan.

## II) Theoretical Foundations

Theoretical foundations typically aim for parsimony, establishing a functional framework with fixed and small bundles of explanatory variables to ensure a well-specified model (Heinze et al., 2018). However, this rigidity poses critical challenges when applied to complex social phenomena like government spending on pension and welfare systems.

First, models constituted solely on directly measurable parameters suffer from the omitted variables problem in social science studies, leading to model misspecification and non-normally distributed error terms (Busenbark et al., 2022). Second, such rigid theoretical platforms inherently assume homogeneity, which prevents them from capturing the fundamental heterogeneity and unique behavioural constraints—like divergent consumer savings or utility functions—between distinct countries, such as China and Western economies. While complex techniques exist to address heterogeneity (e.g., using models like PMG or penalized regressions), they often introduce mathematically challenging penalty terms that lack clear economic interpretation.

Given these limitations, relying exclusively on a rigid theoretical model for variable selection is unsuitable for this investigation. This study, therefore, adopts a theoretical-empirical hybrid model (Talbot & Massamba, 2019). This approach is partially validated by core economic theory but is primarily supported by the empirical findings of well-cited previous research. As a result, influential factors are chosen from a broad pool of variables demonstrated to be relevant in the field, ensuring the model's relevance to the specific socio-economic context of China.

To confirm the reliability of these selected parameters, the study addresses the crucial distinction between correlation and causality. A repeated finding of correlation between variables X and Y is insufficient if that relationship is merely spurious due to shared stochastic trends in the time series (Engle & Granger, 1987). Consequently, our empirical approach is built around the criterion of cointegration. This long-run interconnection not only establishes a meaningful equilibrium but also strongly indicates a genuine causal linkage (Granger, 1988), thereby validating the chosen variables as true long-term determinants of Chinese government pension expenditure.

China's pension system is essentially a governmental fund with two massive sub-components (urban and rural funds), leading to substantial restrictions on its investment portfolio. The fund is generally prohibited from investing its resources in the Chinese stock market, being largely limited to national bank deposits and federal government bonds (Wenpei, 2016). Given the long-run interest rate, the overall inflation-adjusted return on these conservative investments could, therefore, be less than zero. Table 1, compiled from the World Bank website, illustrates the net returns of China's pension fund derived from its bank deposits.

**Table 1.** *Net Returns of Chinese Banks Deposit Interest*

| Year | Inflation Rate | Nominal Deposit Rate | Real Deposit Rate |
|---|---|---|---|
| 2012 | 2.6 | 3.0 | 0.4 |
| 2013 | 2.6 | 3.0 | 0.4 |
| 2014 | 1.9 | 2.8 | 0.9 |
| 2015 | 1.4 | 1.5 | 0.1 |
| 2016 | 2.0 | 1.5 | -0.5 |
| 2017 | 1.6 | 1.5 | -0.1 |
| 2018 | 2.1 | 1.5 | -0.6 |
| 2019 | 2.9 | 1.5 | -1.4 |
| Average | 2.14 | 2.04 | -0.1 |

All the data are reported in percentage.

As shown in Table 1, the average real deposit interest rate, the inflation-adjusted nominal rate, was negative, indicating that the pension fund's investment performance from 2012 to 2019 was associated with a loss. Considering the performance of government bonds, the best-case scenario is that China's pension fund investment return amounts to little more than the net benefits of inflation-indexed bonds. Consequently, the investment returns have negligible impact on the fund's in-flow capital. In this situation, the fund's financial sustainability is solely dependent on maintaining the balance between the number of contributors and pensionaries.

To illuminate the fundamental basis of the contributor-pensionary balance, the next step involves analyzing the demographic structure of the population. Employing a quantitative index will better illustrate the gap between the current situation and the optimal equilibrium. To this end, an analysis of the dependency ratio is beneficial. As defined by the UN (2006), the ratio indicates how many seniors and children must be supported for every 100 persons of working age. The specific formula to calculate the dependency ratio is as follows:

$$Dependency\ Ratio = \frac{Pop(0-14) + Pop(65+)}{Pop(15-64)} \times 100 \qquad (1)$$

Where $Pop(.)$ represents the number of persons within the time interval indicated in the subscript.

While this formula is the principal method to calculate the dependency ratio, some studies utilize a time interval of [0 - 17] for the underaged population and 60+ for economically inactive persons. However, this study will adhere to the reference formula above. The data illustrating the ratio movements, the number of births, and the median life expectancy for China from 2012 to 2019 are provided in Table 2.

**Table 2.** *China's Demographic Data*

| Year | Dependency Ratio | Birth per 1,000 people | Life Expectancy |
|---|---|---|---|
| 2012 | 34.90 | 12.73 | 74.85 |
| 2013 | 35.30 | 12.71 | 75.14 |
| 2014 | 36.20 | 12.55 | 75.44 |
| 2015 | 37.00 | 12.40 | 75.73 |
| 2016 | 37.90 | 12.24 | 76.03 |
| 2017 | 39.20 | 12.09 | 76.32 |
| 2018 | 40.40 | 11.93 | 76.62 |
| 2019 | 41.50 | 11.67 | 76.79 |
| Change Over Period | 2.36 | -1.04 | 0.32 |

**Source:** World Bank Database
"Life expectancy" and "Changes Over Period" are measured in years and percentage, respectively.

As observed in Table 2, the dependency ratio demonstrates a strictly increasing trend, reflecting a constant rise in the numerator of the formula. This upward movement is driven by two opposing factors: a downward trend in net births counterbalanced by a gradual increase in life expectancy. Consequently, the growth rate of the senior population surpasses that of underaged children, confirming a trend of population aging (see Figure 1). Critically, the growth rate magnitude of the dependency ratio is approximately seven times greater than the life expectancy increase, which implies a significant decline in the number of working-age people.

As discussed, the dependency ratio captures the interactions among the three main population segments from a labor market perspective. Since no country exists without children and senior citizens, the ratio will never reach zero; instead, it tends toward an optimum minimum. This optimal point is reached when the number of economically inactive people is minimized in favor of the working-age population.

From the perspective of a pension system, the optimal equilibrium where the number of contributors substantially exceeds the number of pensionaries, resulting in capital accumulation, is known in social science literature as the Demographic Dividend (DD). The DD represents a unique period during a country's demographic transition—specifically, the shift from high fertility and low life expectancy toward a lower birth rate and a high senior population (Ross, 2004). In essence, the Demographic Dividend constitutes an exceptional period within a country's entire history.

The Demographic Dividend (DD) corresponds precisely to an optimum range of the dependency ratio. In this phase, the total expenditure required to support the two dependent groups (children and seniors) diminishes, which directly favours an increase in national savings. Consequently, the enhancement of gross capital formation pushes the Gross Domestic Product (GDP) toward upper levels. The economy thus experiences a period of prosperity marked by high income per capita and significant welfare gains (Zaman & Sarker, 2021).

The crucial question, however, is what happens after the DD era. To illustrate, consider a country characterized by a diminishing fertility trend that subsequently faces gradual population aging. In this scenario, the dependency ratio would jump significantly, as the newly retired individuals are not sufficiently substituted by the emerging young generation of contributors.

Although the Demographic Dividend (DD) brings extraordinary economic opportunities, it fundamentally violates the natural demographic balance and eventually places the pension system under unbearable pressure. In response to this demographic shift, the government is often compelled to increase expenditure on social security and pension funds. In other words, while the DD temporarily boosts GDP growth for a restricted time window (typically one or two decades), it ultimately depresses the overall economy by necessitating increased budget deficits. For example, by analyzing China's labor market, Meng (2023) confirmed that the country successfully enjoyed the benefits of the DD from 1975 to 2015. However, the subsequent, significant downward trend in labor market supply post-2015 clearly signals a change in this favorable demographic equilibrium.

The combination of increasing government expenditures and a steady, gradual decline in economic growth strongly supports the idea that China has lost its demographic advantage, signaling that the economy must prepare for the resulting pressures. While this issue can be managed in the long run, the only short-term solution would be to import ready-to-work labor forces from other countries. Unfortunately, this is not a feasible plan for China; given the country's massive population, the labor market gap is simply too large to be filled solely by foreign workers.

Furthermore, China is not socially prepared to welcome such a large influx of outsiders. Due to its unique culture and heritage, China maintains a specific conservative community (Fang, 2009). Additionally, the modern Chinese language (Mandarin), with its unique symbolic alphabet, is far more complicated to learn than languages like English. In sum, China's society is not only unready to absorb a vast foreign population, but its cultural and linguistic complexities have also significantly reduced its attractiveness to international mobile labor forces (Chu, 2021).

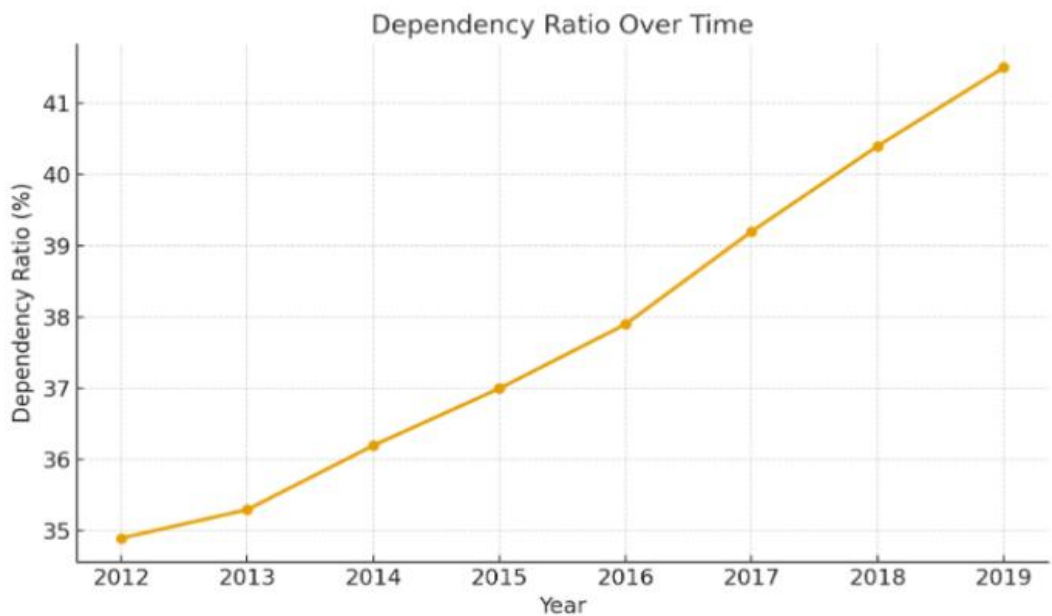

**Figure 1.** Dependency ratio evolution 2012-2019

As established in the introduction, the focus for pension fund solvency must be on maximizing the contribution period of the early young generation, the labor forces with the longest potential working life. However, increasing the employment rate within the 15-24 age bundle requires not only proportional labor market demand but is also inversely correlated with the average age of first-time entrants. Unfortunately, there is a fast-growing tendency among young people to continue their education past high school, especially in developing countries that heavily subsidize higher education.

Therefore, a policy that effectively redirects individuals away from university enrollment and toward the labor market could significantly bolster the pension system's sustainability by increasing the number of long-term contributors.

Research has consistently explored the impact of university tuition fees on higher education enrolment. For instance, Hemelt & Marcotte (2011) studied bachelor's degree students in the U.S. and found that an increment of approximately $400 in university fees led to about a one percent decline in the total number of enrolled students. Sa (2014) reached similar conclusions in the U.K. These findings suggest that higher education costs can serve as a deterrent for potential students, pushing some toward the labour market.

A novel investigation by Alecke et al. (2013) analysed the effect of Germany's 2005 Federal Constitutional decision, which mandated seven out of sixteen federal states to establish a flat tuition rate for new university registrants. Their results demonstrated an asymmetric influence based on gender: the new rule increased the desire of young male students to migrate to the nearest state offering a free educational system, while the impact on females was not significant. This indicates that the tuition fee policy, while not substantially reducing the overall number of enrolled students, promoted in-land immigration, which consequently enhanced the total consumption and provided a pool of well-educated labour

Another crucial element affecting the pension system is the fertility rate, as a reasonable birth rate is essential for stabilizing the labor market supply. Conversely, a sharp decline in the fertility rate, as observed in China, can reduce employment rates and cause financial disequilibrium in the pension system. Peter et al. (2020) analyzed China's demographic changes, focusing on the childbearing women population. They demonstrated that while terminating the one-child policy may boost the fertility rate, the simultaneous downward movement of the childbearing women population negates this effect. Consequently, the policy is unlikely to hit its target unless accompanied by fundamental economic-demographic reforms.

In an interesting discussion, Zhou et al. (2023) addressed the Demographic Dividend (DD) phenomenon in China through the lens of the average university educational level. They concluded that the size of the educated population has a direct and positive correlation with the DD. Furthermore, they argued that the DD relies on both the overall population and labour forces; consequently, a massive number of ready-to-work people, without a well-developed labour market and sufficient labour supply, will not allow the country to fully enjoy the DD benefits.

Prior to this, Cai (2020), in an analysis of China's extended demographic transition, stated that the period of a prosperous and cheap young labour force has definitively ended, and the country no longer takes advantage of demographic dividends. Nevertheless, Cai posited that by executing some primary reforms, it is possible to increase senior people's employability to enhance their lifestyle and slightly reduce the public pension financial imbalance. Ultimately, he concluded that a combination of increasing the retirement age, implementing labour market reform in favour of older workers' empowerment, and promoting the population of childbearing women could potentially achieve another wave of the demographic echo effect.

The focus now shifts to the expenditure side. Just as pension funds seek an extended period of employment, they also favor a short life interval following retirement. Consequently, population aging is a universal concern among all pension systems. In the context of China, Liao et al. (2020) employed a Markov model to analyze the basic pension system and concluded that while the termination of the one-child policy might provide some short-run benefits, it ultimately harms the fund's long-run sustainability. Furthermore, population aging remains the primary factor globally that threatens the financial sustainability of pension systems.

Wang et al. (2019) investigated the Chinese pension fund through a novel perspective, utilizing a pioneering approach to project and analyse the feasible scenarios and feedback of the system from 2019 to 2070. Their analysis forcefully argued that continuing the current situation will lead to a severe financial crisis. Moreover, they demonstrated that if the retirement age were increased to 65 years and all childbearing women were to give birth to at least two children, the budget deficit could be gradually reduced by half, though this benefit would materialize no earlier than 2040.

Subsequently, Liu & Zhao (2023) studied the impact of population aging on China's fiscal sustainability by decomposing the entire territory into four disjoint districts. They inferred that sustainability is highly fragile in districts with elevated population density. Consequently, they proposed that in-land migrations—from industrialized provinces toward the country's least developed partitions—would help boost the fertility rate, decrease citizens' average age, and provide better fiscal sustainability, particularly concerning government public expenditures. These deductions are further validated by the consistent findings of Kudrna et al. (2022) and Meier & Werding (2010).

Urbanization, a phenomenon that accelerated after the 2008 financial crisis, has fundamentally restructured traditional households where one income earner supports the entire unit. Specifically, urbanization has reduced the average household size, leading to a substantial increase in the number of single-person households. This shift initially presents a benefit to pension systems because more individuals are economically compelled to work compared to traditional patterns, thus increasing the number of reachable contributors. Nevertheless, this process also resonates with the simultaneous falling trends of fertility rates. Consequently, a significant nexus exists between urbanization and the financial sustainability of pension schemes.

It could be claimed that the correlation between urbanization and sustainability is established indirectly through economic growth channels. To analyse this, recall that throughout this study, government expenditures on the public sector are selected as the proxy for the financial sustainability of the pension system. While unbounded growth in these expenditures would lead to pension fund bankruptcy, measuring these expenses as a share of GDP provides a unique and necessary comparative tool to analyse the cost of the pension system relative to the overall size of the economy. The trend of urbanization in mainland China is plotted in Figure 2.

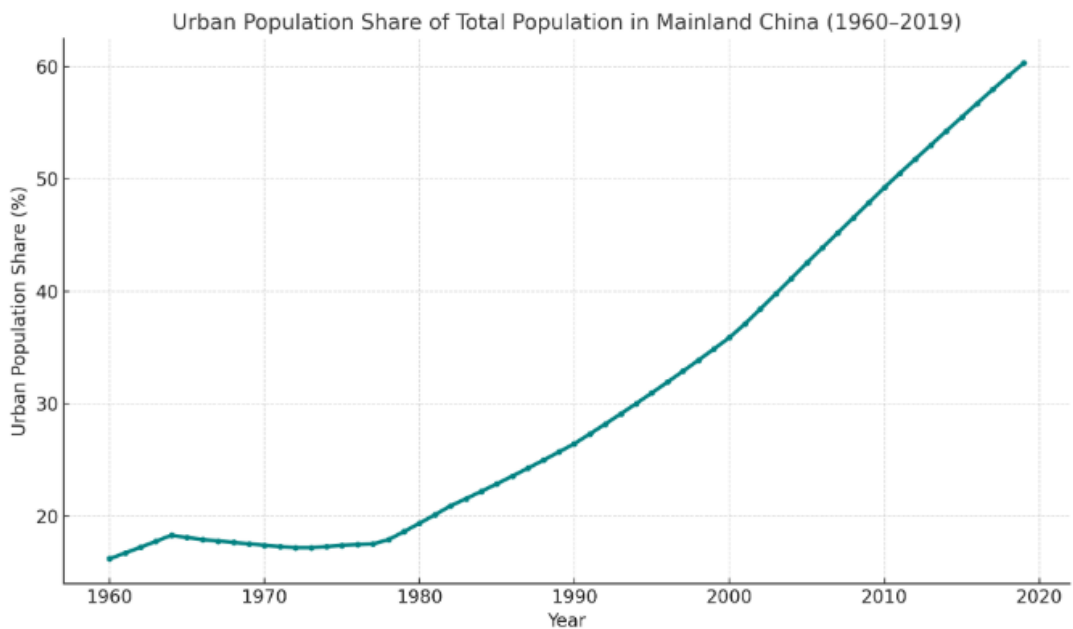

**Figure 2**. Urbanization movement through 60 years.

**Source:** World Bank Database

To illustrate, suppose last year's pension expenditure was equivalent to 10 percent of total GDP. At the end of the current year, assumed economic growth reached 5 percent, while the increment in public expenditure increased slightly by only 3 percent. Assuming the government did not simultaneously restrict the liability column of the pension fund's balance sheet, the expenditure share at the end of the current year will have decreased to approximately 9.8 percent of total GDP. Since the ratio of paid expenses has declined compared to the previous year, the government can transfer this positive difference to the asset column of the fund's balance sheet, stabilizing its cash inflow-outflow and advancing it toward a sustainable equilibrium. Therefore, the analysis will first investigate the nexus between urbanization and GDP and then shift focus to the correlation between economic growth and the financial sustainability of the public pension system.

Increasing the share of people living in urban areas provides a unique opportunity for local industries to access a high supply of cheap labour. This occurs because agricultural labour forces, primarily residing in rural districts, migrate to urban centres. This process offers a distinct advantage to firms that rely heavily on labour forces rather than technological advancement. Consequently, within the framework of a Cobb-Douglas production function, productivity gains are often achieved by substituting capital with more workers.

Furthermore, the employment of this inexpensive labour force leads to a decline in marginal costs, establishing a significant price advantage for the industrial sector. This cost-efficiency can, in turn, drive the sector toward higher production levels, provided there are no bottlenecks in the demand side and firms can freely and easily export their excess production.

The relationship between economic growth and urbanization has been extensively studied. Chen et al. (2014) analysed this nexus using a panel data investigation spanning 226 countries and territories from 1980 to 2011. They concluded that a strong correlation exists between these two parameters. However, their outcomes did not validate a direct causality between the urbanization growth rate and GDP levels.

Further complicating the relationship, Nguyen & Nguyen (2018), in their seminal paper, investigated the interconnection between economic growth and urbanization across a panel of ASEAN countries from 1993 to 2014. Utilizing static and dynamic models estimated with D-GMM and PMG techniques, their results revealed a two-regime nonlinear specification. The regimes were separated by a threshold value for the urbanization level, estimated at approximately 70 percent for both models. This implies that urbanization positively impacts GDP up to this threshold but exerts a negative influence on economic growth once that level is exceeded.

In contrast to the conclusion of a threshold effect, Hofmann & Wan (2013), utilizing a simple OLS panel estimation, found that the interconnection between GDP and urbanization is positively linear and explicitly ruled out the existence of any threshold effect. Introducing further complexity, Henderson (2003), writing more than ten years earlier and in a reverse direction from all the aforementioned studies, argued that no statistically significant relationship exists between these variables, whether linear or nonlinear.

The concentration now shifts to the nexus between economic growth (GDP) and the public pension system. As illustrated in the earlier hypothetical example, if the GDP growth rate exceeds the rate of increase in pension expenditures, the government gains the capacity to stabilize the fund's balance sheet by injecting excess financial resources compared to the previous period. Conversely, increasing pension payments and providing allowances for older individuals who have not contributed to the system can severely harm economic growth rates by rapidly escalating public expenditures, as Temsumrit (2023) demonstrated in the context of the Thailand pension system.

In another interesting line of research, Bijlsma et al. (2014) studied the effect of pension systems' savings on the economic growth rate. They stated that the excess revenues of a sustainable pension fund can be used for investment goals, which directly enhances gross capital formation, the main parameter driving GDP promotion. Consequently, a pension system with a substantial budget deficit could severely harm the economy's growth trajectory by necessitating increased government public expenditures. Moreover, if that deficit is financed through bank loans, it pushes the economy toward a tightening monetary policy by increasing the nominal interest rate. In sum, it can be claimed that GDP growth is positively correlated with the pension system's financial sustainability.

Some ideas assert that the pension system should be governed by the private sector. However, consider the situation where a private pension fund goes bankrupt: contributors would lose a substantial portion of their expected future cash flow. As citizens, they would justifiably expect the government to fill this financial gap. The government is, in effect, compelled to intervene because ignoring such a crisis would certainly fuel public dissatisfaction, leading to national protests and riots. Consequently, the incumbent government, motivated by the desire for re-election, would be forced to rapidly provide financial resources, likely by increasing budget deficits.

To prevent such a vicious scenario in the first place, all workers must become contributors to the government basic pension fund. Employees who desire more extensive financial coverage are, of course, free to secure those excessive benefits through plans offered by the private sector. However, every market participant should enjoy a basic retirement plan that is guaranteed and managed by the government. As a result of this foundational guarantee, the sustainability of the overall pension system would have a negligible impact on the total economic sphere, especially on GDP, which is the usual proxy for overall economic conditions.

### III) Data and Variables

The government expenditure on public pensions and social welfare, as a percentage of China's GDP, is the main proxy for China's pension expenses. Unfortunately, there exists only an annual time series from 1995 to 2006[1] for the government's expenditure in RMB and another time series for 2011 to 2020[2] for the share of expenditure of total GDP. As a result, there is no data for 2007 to 2010. However, a complete time series was needed to be able to estimate a model, and hence, the missing data have been interpolated using a weighted moving average technique[3]. To give an illustration of the time series overall trend, the available data for a period of 25 years has been plotted in Figure 3 as follows:

---

[1] https://www.ceicdata.com/en/china/government-expenditure-by-other-category/government-expenditure-pension-and-social-welfare-ps
[2] https://www.statista.com/statistics/251650/public-pension-expenditure-in-china-as-a-share-of-gdp/
[3] For a practical example of weighted moving average interpolation, please see https://real-statistics.com/time-series-analysis/stochastic-processes/handling-missing-time-series-data/

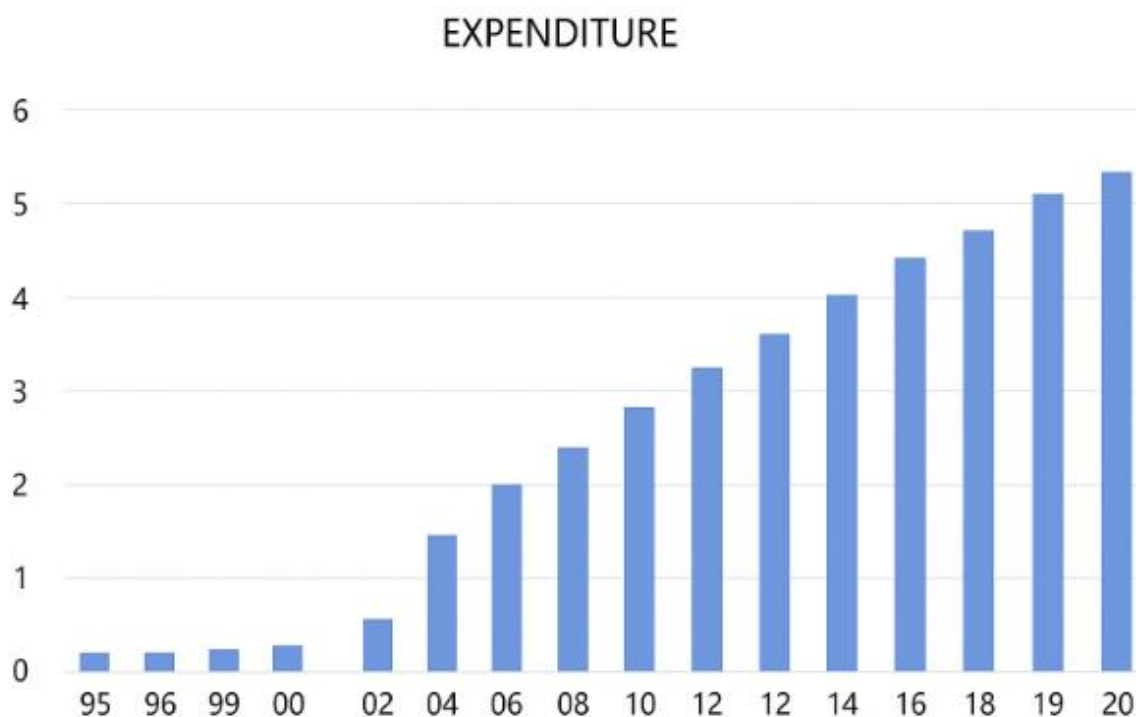

**Figure 3.** Government's Expenditures on Public Pension and Welfare

As can be deduced, after 2006, the series representing public pension expenditures experienced a massive jump, rising from less than one percent to over five percent by 2019. This indicates that pension expenditure has grown at a high-speed rate. However, this illustrates only half of the problem. The other part concerns the share of pension fund assets in relation to China's total GDP. Data for this second time series is available from 2011 to 2020[1].

An interesting finding is that the average growth rate of government expenditure on public pensions during 2012–2020 was approximately 8.1 percent, while the growth rate for pension fund assets was significantly higher at 13.03 percent. This finding suggests that the China pension fund is operating in alignment with Samuelson's (1958) sustainability criterion. Therefore, based on this inference, and in contrast to most well-developed countries that must resort to increasing the retirement age to keep their pension systems solvent, China is currently on the correct path toward financial equilibrium.

According to the extended discussions provided in both the introduction and the theoretical foundations' parts, the exogenous variables include birth rate in 1000 people (Birth[2]), urbanization as the share of people living in urban areas of the total population (Urban[3]), economic growth rate as a proxy for macroeconomic conditions (GDP[4]), the participation rate of youth in the age range of 15 to 24 (Employment[1]), and the percentage of senior people with age higher than 65 of total population (Oldies[2]).

---

[1] https://tradingeconomics.com/china/pension-fund-assets-to-gdp-percent-wb-data.html
[2] https://data.worldbank.org/indicator/SP.DYN.CBRT.IN?end=2019&locations=CN&start=2011
[3] https://ourworldindata.org/urbanization
[4] https://www.macrotrends.net/countries/CHN/china/gdp-growth-rate

**IV) Modeling and Statistical Inference**

Following standard econometric practice, the analysis begins by estimating correlation coefficients. This initial step is vital because it offers a preliminary approximation of the model's coefficients and highlights co-movements among the variables, serving as an initial check for cointegration. We utilized the Pearson correlation coefficient, a ubiquitous statistical tool in economics and social sciences. The full results are presented in Table 3 as a lower triangle matrix, reflecting the measure's inherent symmetry.

**Table 3.** *Correlation Matrix*

|             | Expenditure | Birth | GDP   | Employment | Oldies | Urban |
|-------------|-------------|-------|-------|------------|--------|-------|
| Expenditure | 1           |       |       |            |        |       |
| Birth       | -0.72       | 1     |       |            |        |       |
| GDP         | -0.63       | 0.28  | 1     |            |        |       |
| Employment  | -0.93       | 0.91  | 0.44  | 1          |        |       |
| Oldies      | 0.96        | -0.87 | -0.52 | -0.97      | 1      |       |
| Urban       | 0.94        | -0.89 | -0.44 | -0.99      | 0.98   | 1     |

The results reveal that birth rate, economic growth, and the youth employment rate (ages 15–24) all negatively influence government expenditure on public pensions, thus contributing to pension system sustainability. As anticipated, the coefficient for the share of old people is positive. However, a novel and interesting finding is that urbanization is also co-directed with increasing pension expenditures. In terms of magnitude, the share of old people has the largest absolute coefficient, indicating that an increase in this ratio will fundamentally drive the growth of government expenditure. Conversely, while GDP growth rate decreases total expenditures, its coefficient has the smallest absolute value. This divergence suggests that pension system sustainability appears to be more significantly dependent on social parameters than on economic indicators.

The next crucial step in the modeling process is categorizing variables based on their stochastic trends, commonly known as unit roots. Identifying and accounting for this feature is mandatory, as ignoring it can lead to spurious regression, a phenomenon first described by Granger and Newbold (1974). A time series with a unit root implies that a random shock will permanently alter the series' trajectory, preventing it from returning to its long-run average (Raeisi Sarkandiz & Ghayekhloo, 2024). Numerous statistical tests exist to examine the presence of a unit root or the covariance stationarity of a time series, but their methodological differences primarily stem from the definition of the null hypothesis and the resulting distribution of the test statistics.

The primary difference among unit root tests lies in their null hypothesis. For example, the null hypothesis for the Augmented Dickey-Fuller (ADF) test (Said & Dickey, 1984) and the Phillips-Perron (PP) test (Phillips & Perron, 1988) assumes the presence of a unit root, whereas the Kwiatkowski et al. (KPSS) test (Kwiatkowski et al., 1992) asserts the stationarity of the time series. Although ADF and PP are the most commonly applied tests in empirical studies, the moving average (MA) component of a time series can significantly impact their power, often leading to divergent results.

---

[1] https://www.macrotrends.net/countries/CHN/china/labor-force-participation-rate
[2] https://fred.stlouisfed.org/series/SPPOP65UPTOZSCHN

Ryan and Giles (1998) found that the PP test is more powerful in the presence of positive MA elements, while ADF exhibits lower Type I errors when dealing with negative MA components. Crucially, previous research by Dods and Giles (1995) postulated that real-world variables typically exhibit negative moving average behaviour. Consequently, the present study relies on the ADF test to ensure a more robust analysis given the nature of macroeconomic data. The test assumes the following specification for the time series:

$$\Delta y_t = \alpha + \beta t + \gamma y_{t-1} + \sum_{i=1}^{p-1} \delta_i \Delta y_{t-i} + \varepsilon_t \qquad (2)$$

Where $\alpha$ is the intercept, $\beta$ is the coefficient of deterministic time trend, $\varepsilon_t$ is the $i.i.d$ error terms with variance $\sigma^2$ and zero expected value, and $p$ represents the optimum number of lags which will be selected using information criterion. The model conventionally estimates by the ordinary least squares (OLS) technique; hence it is assumed that the error terms are distributed normally. The test inference carries on the estimated value of $\gamma$ coefficient using standard t statistics specified as:

$$DF_\tau = \frac{\hat{\gamma}}{SE(\hat{\gamma})} \qquad s.t. \quad \begin{cases} H_0: \gamma = 0 \\ H_1: \gamma < 0 \end{cases} \qquad (3)$$

It is critical to note that the test statistic does not follow the standard Student's t distribution under the null hypothesis; therefore, its finite-sample critical values were extracted from the results published by Cheung and Lai (1995). The complete test outcomes for the selected database are reported in Table 4.

**Table 4.** *ADF Unit Root Test*

| Null Hypothesis: The Time Series Has a Unit Root |||
|---|---|---|
| Lag Selection: SIC |||
| Variable | ADF-Stat. | P-Value |
| Employment | -3.27 | 0.03 |
| Urban | -1.72 | 0.41 |
| Birth | -0.69 | 0.83 |
| Expenditure | -0.10 | 0.94 |
| GDP | -1.38 | 0.57 |
| Oldies | 1.32 | 0.99 |

The results indicate that all variables, except for the employment rate, possess at least one unit root at the five percent significance level. This outcome is promising, as it substantially raises the probability of a cointegrated vector being present. While the rejection of the null hypothesis for the employment rate does not definitively imply covariance stationarity—given the possibility of a near or fractional unit root (Baillie, 1996)—the existence of a mean-reverting or long-memory trend in this single variable will not impact the overall cointegration structure of the system. Therefore, no further investigation into the stationarity of the employment rate is required.

The analysis now proceeds to the next step: estimating a linear model using the following specification:

$$Expenditure_t = C + Birth_t + GDP_t + Employment_t + Oldies_t + Urban_t + \varepsilon_t \quad (4)$$

Where $\varepsilon_t$ represents the error terms or residuals. While the initial priority was to estimate a log-linear model to leverage the direct interpretation of coefficients as elasticities, this model—even in a dynamic, ARDL specification—suffered from a violation of the principal assumptions for Ordinary Least Squares (OLS) and Maximum Likelihood (ML) estimators. Specifically, the log-linear form exhibited significant autocorrelation and heteroscedasticity in the error terms. Although employing a Heteroscedasticity-Autocorrelation Consistent (HAC) covariance matrix (Newey & West, 1987) successfully mitigated these two issues, this fix unfortunately introduced a new problem: non-normality of the innovations.

It is well-known that in the presence of non-normally distributed residuals, the critical values for $F$ and $t$-tests do not follow the standard distribution, rendering their outcomes unreliable (Raeisi Sarkandiz & Bahlouli, 2019). Even though information criteria (AIC and BIC) suggested an ARDL (1,0,1,1,0,0) model, the fundamental diagnostic issue was the lack of cointegration in the dynamic specification. This meant that the estimated dynamic model constituted a spurious regression, making it useless for statistical inference. Therefore, the only viable alternative was to revert to the variables in their raw (non-logarithmic) format for estimation. The estimated linear model is reported in Table 5 as follows:

**Table 5.** *Estimated Model*

| Dependent Variable: Expenditure | | | |
|---|---|---|---|
| Estimation Method: OLS | | | |
| Variable | Coef. | t-Stat. | P-Value |
| C | 12.76 | 0.62 | 0.54 |
| Birth | 1.04 | 5.59 | 0.00 |
| GDP | -0.11 | -3.63 | 0.00 |
| Employment | -0.38 | -1.55 | 0.14 |
| Oldies | 0.64 | 4.94 | 0.00 |
| Urban | -0.13 | -0.66 | 0.52 |
| R-Sq = 0.995 | | Adj R-Sq = 0.993 | |
| F-Stat. = 700.60 | | F-Prob. = 0.00 | |
| SIC = -0.40 | | D-W Stat. = 1.68 | |

As presented in Table 5, the estimated parameters are highly promising. The model exhibits a high explanatory power of approximately 99 percent, which is paired with a reasonable Durbin-Watson (D-W) statistic (Durbin & Watson, 1950, 1951). This combination is a crucial indicator of model validity, as Phillips (1986) demonstrated that a high $R^2$ coupled with a near-zero D-W statistic is the chief sign of a spurious regression. Furthermore, the F test—the joint significance test—fully validates the overall model. The logic of the F test is to test the null hypothesis that all estimated coefficients are simultaneously zero against the alternative that at least one coefficient is non-zero. To confirm this joint significance, the F statistic is calculated as follows (Greene, 2002):

$$F - Stat. = \left(\frac{RSS_1 - RSS_2}{P_2 - P_1}\right) \div \left(\frac{RSS_2}{n - P_2}\right) \qquad (5)$$

Where $RSS_i$, $i = 1,2$ are the residuals sum of the square of restricted, all coefficients are equal to zero, and unrestricted models and $P_i$, $i = 1,2$ are the number of estimated parameters in each model. Analysis of the F statistic's structure yields three key facts. First, since it is calculated as the ratio of two variances, the statistic is inherently a non-negative value. Second, the test is based on the null hypothesis that all coefficients in the *restricted* model vanish; consequently, applying the F test implicitly requires the model to be estimated with an intercept. This requirement is precisely why virtually all statistical software packages automatically include an intercept parameter in model estimations.

The third critical fact is that when both denominators incorporate the appropriate degrees of freedom, the F statistic essentially becomes the ratio of two Chi-square distributions. Furthermore, as Box (1953) demonstrated, a $\chi^2$ distribution is derived from the sum of squares of independent standard normal distributions. Therefore, the inherent and crucial presumption underlying the valid use of the F test is that the error terms must be normally distributed. In the subsequent step, an appropriate statistical test will be applied to investigate whether this normality condition is satisfied.

Among the estimated parameters, the employment rate and urbanization are found to be not statistically significant at the five percent level. However, this result should not warrant the removal of these variables. In fact, estimated models without these two elements exhibited lower explanatory power and introduced the issue of non-normal error terms. This lack of significance merely suggests that the relationship between these parameters and the response variable is not entirely linear. For instance, a hybrid functional form—such as a combination of linear, quadratic, or hyperbolic functions—might yield better results, but this would shift the modelling concept from regression to interpolation, which falls outside the scope of this study.

The non-stationary nature of most variables unfortunately made it impossible to utilize Granger's (1969) causality test to determine the direction of the interconnections. Furthermore, the database was insufficiently large to effectively employ the Toda-Yamamoto (T-Y) causality test (Toda & Yamamoto, 1995). The T-Y test requires the estimation of a Vector Autoregressive (VAR) model, which, given the number of variables and the optimum lag length (P=2) suggested by the Schwarz Bayesian Criterion (SBC), faced an over-fitting problem and rendered the T-Y test outcomes unreliable.

An interesting finding is that the estimated signs of the coefficients did not fully align with the initial correlation matrix. This contradiction can be easily resolved by recalling that the Pearson correlation estimator assumes normally distributed variables (Asuero et al., 2006). Furthermore, the fact that the birth rate coefficient is higher than the approximated coefficient of the old people population, with both being positive, effectively confirms the demographic dividend for China within this model.

However, all these inferences are contingent upon the validation of the model. Specifically, the residual time series must not only be normally distributed (non-violation of the F test assumption) but must also be homoscedastic and exhibit no serial correlation. Additionally, the model needs to be integrated, well-specified, and benefit from time-insensitive coefficients. The results from these necessary diagnostic tests are reported in Tables 6 and 7, and Figure 4, respectively.

**Table 6.** *Residuals Diagnosis*

| Stat. | Value | D.F. | P-Value |
|---|---|---|---|
| \multicolumn{4}{c}{Breusch-Pagan Test} | | | |
| \multicolumn{4}{c}{Null Hypothesis: Homoscedasticity} | | | |
| F-Stat. | 2.11 | (5,19) | 0.10 |
| L.M. | 8.94 | 5 | 0.11 |
| \multicolumn{4}{c}{Breusch-Godfrey Test} | | | |
| \multicolumn{4}{c}{Null Hypothesis: No Serial Correlation} | | | |
| F-Stat. | 2.69 | (2,17) | 0.09 |
| L.M. | 6.01 | 2 | 0.05 |
| \multicolumn{4}{c}{Jarque-Bera Test} | | | |
| \multicolumn{4}{c}{Null Hypothesis: Normality} | | | |
| \multicolumn{4}{c}{Stat. = 1.43   P-Value = 0.49} | | | |

**Table 7.** *Model Diagnosis*

Residuals Stationary/Unit Root Test
Lag Selection: SIC

| Test | Value | Critical Value | |
|---|---|---|---|
| ADF | -5.07 | -3.00 | |
| KPSS | 0.13 | 0.46 | |

Ramsey RESET Test
Null Hypothesis: The Model is Well-Specified

| Stat. | Value | D.F. | P-Value |
|---|---|---|---|
| t-Stat. | 1.29 | 18 | 0.21 |
| F-Stat. | 1.68 | (1,18) | 0.21 |
| L.R. | 2.23 | 1 | 0.13 |

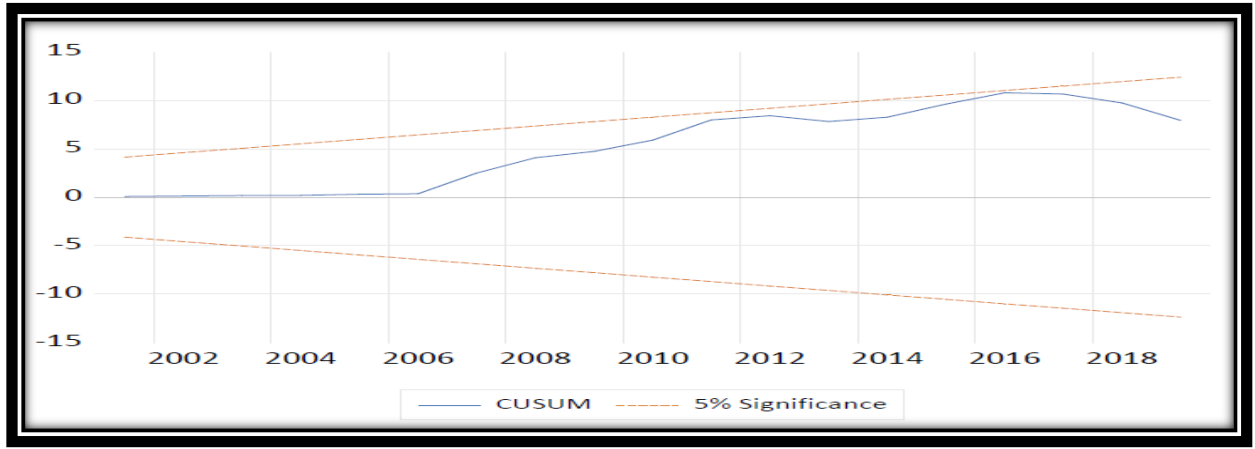

**Figure 4.** CUSUM Stability Test of Estimated Coefficients

According to Table 6, the Breusch-Pagan (1979) test confirmed that the residuals are homoscedastic, and the Breusch (1978) and Godfrey (1978) tests validated that the innovations are not serially correlated. This examination is crucial because if the variance of the error terms were time-dependent (heteroscedastic), the diagonal elements of the variance-covariance matrix would not be equal, invalidating the OLS standard errors.

The subsequent step is to examine the distribution of the residuals. For this purpose, the Jarque-Bera (1980) test, with its null hypothesis of normality, is employed. It is important to note, however, that the rejection of the null hypothesis provides no specific information regarding the underlying data distribution, as the distribution under the alternative hypothesis is unspecified. The test statistic is calculated as follows:

$$JB = \frac{n}{6} \cdot \left( S^2 + \frac{(K-3)^2}{4} \right) \quad (6)$$

Where $S$ and $K$ are sample skewness and kurtosis, respectively. The Jarque-Bera (JB) statistic asymptotically follows a Chi-square distribution with two degrees of freedom (Kim, 2016). It is important to mention that although several alternative normality tests exist, Thadewald and Büning (2004), using a Monte Carlo simulation, demonstrated that the JB test is the optimal choice for relatively small samples, a condition that aligns precisely with this study's dataset. According to Table 6, the calculated test statistic does not exceed the critical value, leading to the crucial conclusion that the null hypothesis of normality is not rejected.

With the non-rejection of the normality null hypothesis, all fundamental assumptions of the OLS technique and the associated significance tests (F and t) have been satisfied. The analysis now moves to the next stage: confirming that the model formulation is well-specified. This involves assessing whether the linear model is the appropriate choice, or if some form of non-linearity should be introduced. For this critical examination, we employ Ramsey's (1969) Regression Specification Error Test (RESET).

Although there are alternative tests, such as those by White (1989) and Teräsvirta et al. (1993), they were deemed unsuitable. Both alternatives consider nonlinearity through the hidden layers of an Artificial Neural Network (ANN), which represents a fundamentally different, non-parametric approach compared to the parametric framework of the present research.

Furthermore, while simulation studies by Prabowo et al. (2020) concluded that the RESET test has higher power than White's but lower than Teräsvirta's, Teräsvirta's statistic is highly sensitive to outliers, unlike the RESET test. Consequently, the RESET test represents the most suitable and robust choice. Its logic involves constructing a restricted model and comparing it to the unrestricted model using an F test to determine if the model is well-specified. The procedure for calculating the RESET test statistic and its hypotheses is described as follows:

The analysis begins by estimating the main model using the Ordinary Least Squares (OLS) method. Subsequently, the fitted values ($\hat{y}_t$) time series according to the original model should be calculated. The unrestricted model is then constructed by incorporating powers of the fitted values ($\hat{y}_t^2, \hat{y}_t^3, ...$) into the explanatory variables' matrix. In the next step, the test statistic is constructed with respect to formula (5), and the null hypothesis considers the coefficients of these newly added variables equal to zero. The calculated statistic follows an F distribution, where $k$ represents the number of independent variables in the original model.

Consequently, the rejection of the null hypothesis would imply possible non-linearity in the model, arguing against a linear construction. However, the outcomes of the test, as reported in Table 7, display that the RESET's null hypothesis cannot be rejected. This finding is crucial, as it indicates that the model is well-specified and confirms that the linear regression form is an appropriate choice for this analysis.

The initial part of Table 7 presents the results of the cointegration analysis. The existence of a cointegration relationship, or, in the sense of linear algebra, a cointegrated vector, guarantees a long-run co-movement among the model's parameters. This implies that while variables may be deviated or even orthogonal in the short run, they exhibit a strong, common trend in value and sign over the long run (Rahman & Raeisi Sarkandiz, 2023). The chosen approach here is the Engle-Granger two-step cointegration test, which first estimates the linear model and then checks the stability (stationarity) of the residual time series in the second step.

While the Johansen (1991) rank test is a powerful alternative, its foundation in a Vector Autoregressive (VAR) model structure, similar to the T-Y test, made its utilization infeasible due to sample size constraints. Furthermore, although the Bounds test of Pesaran et al. (2001) is another widely used cointegration test, it is only applicable to dynamic ARDL models, which is not the specification adopted in this study. The Engle-Granger method was therefore selected as the most appropriate and robust test given the final model's form and data limitations.

Since the aim is to validate the stationarity of the error terms, a hybrid testing method is employed. This approach is necessary because the rejection of the Augmented Dickey-Fuller (ADF) test's null hypothesis (unit root) does not definitively confirm stability. Time series stationarity is only concluded when the ADF test rejects the existence of a unit root *and* the KPSS test (Kwiatkowski et al.) does not reject its null hypothesis (stationarity) (Raeisi Sarkandiz, 2020). The outcomes from Table 7 are decisive: the ADF null is rejected and, simultaneously, the KPSS null is not rejected. Therefore, the residuals follow a stationary trend, which, by definition, means the model is integrated (cointegrated).

The proven cointegration confirms that the variables share a long-run relationship, validating the regression model (4) as genuine and not spurious. Furthermore, as discussed in the theoretical foundations section, cointegration inherently implies causality, although the reverse may be incorrect. This establishes a key finding: the estimated model is causal, and there exists a uni-directional causality from the explanatory variables toward the response parameter.

The analysis has successfully confirmed that the estimated coefficients are unbiased and consistent, the error term variance is optimum, and all significance tests ($F$ and $t$) are reliable. Furthermore, the residuals satisfy all classical assumptions: they are homoscedastic, normally distributed, and exhibit no serial correlation. Crucially, the model has been confirmed as correctly specified as a linear regression. Given the cointegration validation, a long-term relationship exists among the variables, confirming the regression is genuine (not spurious).

However, the analysis has not yet determined if the estimated parameters are stable throughout time. While the variables' long-run co-movement is established, the individual coefficients could still exhibit time-varying behaviour. This is a critical aspect of econometric modelling because long-run policy implications based on non-stable (or time-varying) factors could be fundamentally illogical and misleading. Therefore, the final step requires investigating the temporal stability of the estimated parameters.

In this regard, and in the final diagnostic step, the time-insensitivity (or stability) of the estimated coefficients is tested by employing the Cumulative Sum (CUSUM) method. The central idea of this test is to determine if the estimated coefficients of the primary model remain equal across a rolling-window estimation procedure. This robust method was initially proposed by Page (1954) and subsequently generalized by Woodward and Goldsmith (1964).

To implement the CUSUM test, the analysis follows the procedure established by Brown et al. (1975). This approach constructs the test statistic based on recursive residuals to monitor deviations in the coefficient estimates over time. To proceed with this formal calculation, suppose:

$$\Psi_t = \frac{1}{\hat{\sigma}} \cdot \sum_{k+1}^{t} \widehat{w}_t \quad s.t. \quad t = k+1, k+2, \ldots, T \quad (7)$$

Be the cumulative sum of the model up to time t, $\hat{\sigma}$ is the estimated standard deviation equal to $\sqrt{\frac{R_T}{T-k}}$ and $R_T$ is the residuals sum of square after estimating the model for all the data. Also, as it is obvious, k is the number of explanatory variables plus one for considering the intercept. In the above formula, the procedure to calculate $\widehat{w}_t$ is as follows:

Let $b_t$ be the OLS estimate of $\beta$ for the first t observations instead of the entire sample. Then if $X_t$ and $Y_t$ be the values of explanatory and dependent variables up to time t in a matrix specification,

So, we have,

$$b_t = (X_t'X_t)^{-1} X_t'Y_t \quad s.t. \quad X_t'X_t \quad is\ non-singular \quad (8)$$

Therefore, we would have:

$$\widehat{w}_t = \frac{y_t - x_t'b_{t-1}}{\sqrt{1 + x_t'(X_{t-1}'X_{t-1})^{-1}x_t}} \quad s.t. \quad r = k+1, \ldots, T \quad (9)$$

Such that $X'_{t-1} = [x_1, \ldots, x_{t-1}]$ and $Y'_t = [y_1, \ldots, y_t]$. As a result, if all the coefficients that estimated using a rolling-window time horizon are equal, then $w_t \sim N(0, \sigma^2)$. Consequently, it implies that:

$$E(\Psi_t) = 0, \quad Var(\Psi_t) = t - k, \quad Cov(\Psi_t, \Psi_s) = \min(t, s) - k \quad (10)$$

Ultimately, if $\Psi_t$ plots against t, the CUSUM curve will be illustrated. As Figure 4 clearly illustrates, there is no intersection between the CUSUM values (the blue line) and the 5% significance bandpass lines (the red lines). This decisive outcome indicates that the coefficients' volatilities remain within a reasonable interval, confirming that the long-run model is stable throughout the entire estimation period.

## V)    Discussion of Findings

The estimated coefficients reveal that GDP positively impacts the sustainability of the Chinese pension system. However, as previously noted, its magnitude is insufficient to fully offset the detrimental effects of other parameters. This finding aligns with the arguments presented by Temsumrit (2023) and Bijlsma et al. (2014). Conversely, the results show that urbanization is positively correlated with system sustainability; however, its estimated coefficient is not statistically significant, suggesting a non-linear nature to this relationship. These conclusions find support in the work of Nguyen and Nguyen (2018) but contradict the statements made by Hofmann and Wan (2013) and Henderson (2003).

The estimated model demonstrates that an increment in the old-age population severely harms sustainability by directly increasing government public expenditure. This finding confirms that population aging is the principal variable threatening the system's stability. This result strongly aligns with the arguments presented by Liu and Zhao (2023) and Liao et al. (2020).

Analysis of the paired coefficients for the birth rate and the senior people's population yields the most fundamental outcome of this study. The sign of both parameters is positive; however, the estimated value for the birth rate is significantly higher than the magnitude of the approximated coefficient for the old people population. This combination provides strong validation that China is still benefiting from a demographic dividend (DD). This crucial conclusion chiefly contradicts the results of Meng (2023) and Cai (2020), who argued that China's DD era has already passed.

According to the Ministry of Education of the People's Republic of China report[1] in 2021, the number of enrolled students seeking a degree program, including undergraduate, postgraduate, master, PhD, and vocational undergraduate at all types of public universities and colleges reached about 41.63 million students while the number was about 3.8 million in 1990. An interesting point is that the number of international students was about 256,000, which counts for less than one percent.

While temporarily increasing the number of university students can assist the labor market in adjusting toward equilibrium by decreasing the immediate labor supply, it paradoxically creates waves of excess supply in the near future. Despite this, our estimated model shows that the coefficient of young people's employment is negative, which directly implies that increasing the variable's level will push pension funds toward sustainability.

---

[1] http://en.moe.gov.cn/documents/statistics/2021/national/202301/t20230104_1038055.html

Consequently, a policy that discourages a large number of young people from immediately seeking a university degree will guide more individuals directly into the job market, thereby increasing the labor supply. Hence, enhancing university tuition fees—a policy analyzed and suggested by Hemelt and Marcotte (2011) in the U.S. and Sa (2014) in the U.K.—could be an effective mechanism for promoting pension sustainability.

## VI) Conclusion and Policy Implications

This paper presents a rigorous analysis of the factors driving Chinese government expenditure on public pensions, providing a comprehensive, long-run perspective previously lacking in the literature. The study's main contributions, derived from a co-integrated socio-economic model, centre on the following key findings: The model utilized a cointegrated empirical approach to establish a genuine long-run relationship and uni-directional causality from the chosen parameters to government expenditure, and the model was confirmed to be stable throughout the entire estimation period. Crucially, the analysis revealed that China is still benefiting from a demographic dividend (DD), which fundamentally contradicts some widely held assumptions. Furthermore, the findings underscored that stabilizing the pension fund is impossible by relying solely on strong economic growth. Instead, the analysis compellingly argues that comprehensive labour market reforms are needed to replace the eventual absence of the demographic dividend.

**Policy Implications for China's Economy**

The policies are tailored exclusively for the Chinese economy and are based on the magnitudes and signs of the estimated model coefficients and the core discussions provided in this study:

I) Aggressive Utilization of the Current Demographic Dividend (DD) Window: The model validates that China is currently enjoying a DD, evidenced by a birth rate coefficient that is significantly higher than the approximated coefficient for the old people population. The government should use this positive momentum as a limited-time opportunity to aggressively build up the pension fund's assets. Since the pension fund's financial sustainability is solely dependent on maintaining the balance between the number of contributors and pensionaries (due to negligible real investment returns), this window must be used for rapid capital accumulation before the demographic advantage fully passes.

II) Targeted Educational Disincentives to Boost Youth Employment: The estimated coefficient for young people's employment (age 15-24) is negative, directly implying that increasing this variable's level will push pension funds toward sustainability. To maximize the number of long-term contributors, the government should implement a policy to modestly enhance university tuition fees. This policy acts as an economic signal, guiding more young individuals directly into the labour market rather than prolonging their education, which temporarily decreases the immediate labour supply.

III) Prioritization of Structural Reform Over GDP Reliance: The analysis demonstrates that while GDP positively impacts pension fund sustainability, its magnitude is insufficient to fully offset the detrimental effects of other parameters, such as the increasing old-age population. This confirms that stabilizing the pension fund is impossible by relying solely on strong economic growth. Therefore, the government must shift its focus from passive reliance on economic growth to actively implementing comprehensive structural reforms in the labour market. These reforms should favour younger worker recruitment and retention, coupled with enhancements in senior people's employability to slightly reduce the public pension's financial imbalance.

**Suggestions for Further Studies**

A) Investigation of Non-Linear Urbanization Effects: The estimated coefficient for urbanization, while positively correlated with sustainability, was found to be not statistically significant in the linear model, strongly suggesting a non-linear nature to this relationship. Further research should employ a non-linear or threshold-based functional form (e.g., an empirical threshold model) to investigate the exact level of urbanization at which its positive impact on the labour market and pension sustainability begins to reverse for the Chinese economy.

B) Formal Causality Testing with Extended Time Series: The non-stationary nature of the variables and the limited size of the available time series precluded the use of powerful methods like the Toda-Yamamoto (T-Y) causality test in this study. Future studies should utilize a more extensive or recent time series dataset to formally test the direction of causality between the key parameters (e.g., GDP, young employment) and government pension expenditure, thereby providing a definitive statistical confirmation that overcomes the data constraints of the current research

**Declaration**

The author declares no conflict of interest.

This research did not receive any funding assistance from governmental, private, or non-for-benefit organizations.